\documentclass[%
 preprint,
preprintnumbers,
nofootinbib,
 amsmath,amssymb,
 aps,
]{revtex4-1}

\usepackage{graphicx}% Include figure files
\usepackage{dcolumn}% Align table columns on decimal point
\usepackage{bm}% bold math
\usepackage{epstopdf}
\usepackage{wrapfig,epsfig}
\usepackage{mathrsfs}
\usepackage{color}
\usepackage[utf8]{inputenc} 

\newcommand{\Slash}[1]{{\ooalign{\hfil/\hfil\crcr$#1$}}}

\begin{document}

\preprint{J-PARC-TH-0120}

\title{Operator relations for
%QCD constraints on 
gravitational form factors of a spin-0 hadron}% Force line breaks with \\

\author{Kazuhiro Tanaka}
 \email{kztanaka@juntendo.ac.jp}
\affiliation{%
 Department of Physics, Juntendo University, Inzai,
  Chiba 270-1695, Japan}
\affiliation{ J-PARC Branch, KEK Theory Center, Institute of
  Particle and Nuclear Studies, KEK, 203-1, Shirakata, Tokai, Ibaraki,
  319-1106, Japan
}%


%\affiliation{
%\affiliation{
%\affiliation{%


\date{\today}% It is always \today, today,

\begin{abstract}
The gravitational form factors for a hadron, the form factors for the hadron matrix element of the QCD energy-momentum tensor, 
not only describe the coupling of the hadron with a graviton,
but also 
serve as unique quantities for
describing the shape inside the hadron reflecting dynamics of quarks and gluons,
such as the internal shear forces acting on the quarks/gluons and their pressure distributions.
We consider the quark contribution to the gravitational form factors for a (pseudo)scalar hadron,
and derive and clarify the relations satisfied by them
as direct consequences of the symmetries and the equations of motion in QCD, and connections to the generalized parton distributions. Our results reveal the connections between the gravitational form factors and the higher-twist 
quark-gluon correlation effects inside the hadrons.
\end{abstract}

\maketitle

%\tableofcontents

%\section{GPDs and gravitational form factors for a spinless hadron}

\section{Introduction}
The gravitational form factors of hadrons have received considerable attention recently~\cite{Burkert:2018bqq,Polyakov:2018guq,Polyakov:2018zvc,Teryaev:2016edw}.
They represent the form factors for the matrix element of the QCD energy-momentum tensor with the one-hadron states receiving a certain momentum transfer~\cite{Pagels:1966zza,Ji:1996ek},
%not only describe the coupling of the hadron with a graviton,
%but also 
and are recognized as playing unique roles in
describing the shape deep inside the hadrons reflecting dynamics of quarks and gluons,
such as the pressure distributions inside the hadrons~\cite{Polyakov:2002yz,Diehl:2003ny,Belitsky:2005qn}.
Although it is impractical to detect those gravitational form factors directly through the coupling of the hadrons with a graviton,
it is now realistic to determine the gravitational form factors for the nucleon based on the behaviors 
of the generalized parton distributions (GPDs)~\cite{Burkert:2018bqq}
obtained by experiments like deeply virtual Compton scattering (DVCS)~\cite{Mueller:1998fv,Ji:1996nm,Radyushkin:1996nd,Goeke:2001tz,Diehl:2003ny,Belitsky:2005qn}, 
deeply virtual meson production~\cite{Collins:1996fb,Goloskokov:2009ia}, meson-induced Drell-Yan 
production~\cite{Berger:2001zn,Goloskokov:2015zsa,Sawada:2016mao}, etc. (For the present status of the experimental data, 
see, e.g., references in \cite{Burkert:2018bqq}.)

The gravitational form factors for spinless particles allow
a simpler theoretical formulation without spin structures compared with those for the nucleon~\cite{Pagels:1966zza}.
Among them, the gravitational form factors of the pion are particularly interesting quantities,
because their behaviors could reflect nontrivial nature as a Nambu-Goldstone boson
and could be compared with the predictions 
of nonperturbative approaches, a variety of  which have been devised and proposed.
%, in particular, for the pion.
% that could provide a test for the models.
Their empirical information was considered to be severely restricted because
the pion target for measuring its GPDs is unavailable. 
Recently, however, it has been demonstrated~\cite{Kumano:2017lhr} that
the behaviors of the gravitational form factors for the pion 
could be extracted
through the determination of the generalized distribution amplitudes (GDAs)~\cite{Diehl:1998dk,Diehl:2000uv,Diehl:2003ny,Kawamura:2013wfa}
using the Belle data on $\gamma^* \gamma  \rightarrow \pi^0 \pi^0 $.
Thus, the investigation of the gravitational form factors for the pion as well as for the nucleon is a hot topic.
%utilizing the crossing relation between the generalized distribution amplitudes (GDAs) 

In view of this, 
%the point which is still unclear
an urgent task from the theory side is to clarify a maximal set of
%which type of 
(exact and approximate) relations satisfied by the gravitational form factors.
%, it is not still clear
The purpose of this paper is to give a contribution in this direction.
We discuss the relations which hold for the gravitational form factors for a spin-0 particle such as a pion.
We clarify how those relations are derived
as direct consequences of constraints (symmetries and equations of motion) in QCD,
applied to
the gauge-invariant operator corresponding to the quark contribution for the energy-momentum tensor.
In our calculation, we retain all the quark-mass effects as well as the terms associated with the hadron mass, and
this allows us to obtain exact results up to twist four.
Those results involve the relations between the gravitational form factors and the matrix elements 
of the higher twist operators in QCD, in particular, those associated with the moments of the twist-three
as well as twist-four GPDs
for the spin-0 hadron. Furthermore, in addition to those exact relations,
approximate relations using an approach based on the light-cone gauge fixing linked with partonic
interpretations in the infinite momentum frame
are obtained and the result suggests that
that the gravitational form factor associated with the so-called ``D-term'' corresponds to
the twist-three quark-gluon interaction effects.
% based on an approach with the light-cone gauge fixing.
Some of our results are immediately extended to the corresponding relations for the  
gravitational form factors of the nucleon.

\section{Energy-momentum tensor and the gravitational form factors}

In this paper, $|p\rangle$ denotes a spin-0 hadron state with the 4-momentum $p$ as $p^2=m_h^2$. 
We consider matrix elements of the energy-momentum tensor $T^{\mu \nu}$ in QCD, 
$\langle p'|T^{\mu \nu} |p\rangle$, for a hadron which may be a scalar meson, or a pseudoscalar meson such as a pion.
We shall keep all the contributions of quark masses as well as the hadron mass $m_h$, and
%, but we will present also the results and
the implications in the chiral limit are also mentioned.
We denote the independent 4-momenta as
\begin{eqnarray}
&&\overline{P}=\frac{p+p'}{2}\ ,\;\;\; \;\;\; \Delta=p'-p\ ,
%\nonumber\\
%&&q=\overline{q}+\frac{1}{2}\Delta\ ,\;\;\; q'=\overline{q}-\frac{1}{2}\Delta\ ,\;\;\; p=\overline{P}-\frac{1}{2}\Delta\ ,\;\;\; 
%p'=\overline{P}+\frac{1}{2}\Delta\ ,
\label{kinema10}
\end{eqnarray}
and we have the relevant invariants, 
\begin{equation}
\Delta^2 = t\ , \;\;\;\;\; \;\;\;\;\;\overline{P}^2= m_h^2 -\frac{t}{4}\ ,
\label{kinema21}
\end{equation}
for the target of mass $m_h$, such that $p^2=p'^2=m_h^2$.
The (Belinfante-improved) energy-momentum tensor
in QCD is given as~\cite{Ji:1996ek} (see also the article by Jackiw in \cite{Treiman:1986ep})
\begin{eqnarray}
T^{\mu \nu}(x)&=&\sum_{q=u,d,s,\ldots}T_q^{\mu \nu}(x)+T_g^{\mu \nu}(x)\ ,
\label{tmn}
\end{eqnarray}
with the quark part for each separate quark flavor as
\begin{equation}
T_q^{\mu \nu}(x)\equiv \frac{1}{2}\bar{q} (x)\gamma^{(\mu} i\overleftrightarrow{D}^{\nu )} q(x)
=\frac{1}{4}\left(\bar{q} (x)\gamma^{\mu} i\overleftrightarrow{D}^{\nu } q(x)+\bar{q} (x)\gamma^{\nu} i\overleftrightarrow{D}^{\mu} q(x)\right)
\ ,
\label{symtmnq}
\end{equation}
where $q(x)$ is the quark field of flavor $q$, $\overleftrightarrow{D}^\mu \equiv -\overleftarrow{D}^\mu+\overrightarrow{D}^\mu$
with $\overrightarrow{D}^\mu=\overrightarrow{\partial}^\mu-igA^\mu$, $\overleftarrow{D}^\mu=\overleftarrow{\partial}^\mu+igA^\mu$ being the covariant derivative, and $(\mu \nu)$ denotes symmetrization with respect to $\mu, \nu$
indices.\footnote{In the RHS of (\ref{symtmnq}), we have also the term, $-g^{\mu \nu} \bar{q} (x) \left( \frac{i}{2}\overleftrightarrow{\Slash{D}}-m \right) q(x)$,
according to the canonical definition of the energy-momentum tensor, but this term vanishes by the use of the equations of motion and is omitted here and in what follows.} 
The gluon part is given as
\begin{eqnarray}
T_g^{\mu \nu}(x)&\equiv&-F_a^{\mu\rho}(x){F_a^\nu}_\rho(x)+\frac{g^{\mu \nu}}{4}F_a^{\lambda\rho}(x){F_a}_{\lambda\rho}(x)\ ,
\label{totalemt}
\end{eqnarray}
with $F^{\mu\nu}=F_a^{\mu\nu}t^a$ being the gluon field strength tensor.\footnote{There is an ambiguity to separating $T^{\mu\nu}$ into the quark and gluon parts.
We consider the gauge-invariant decomposition by Ji~\cite{Ji:1994av} here.}
The matrix element of the quark part of the energy-momentum tensor (\ref{symtmnq}) is parameterized as
\begin{eqnarray}
\!\!\!
\!\!\!
\langle p'|T_{q}^{\mu \nu}|p\rangle \equiv\langle p'|T_{q}^{\mu \nu}(x=0)|p\rangle 
=
\frac{1}{2}{\Theta}_{1q}(t)\left(tg^{\mu\nu}- \Delta^\mu \Delta^\nu\right)+ \frac{1}{2}{\Theta}_{2q}(t) \overline{P}^\mu\overline{P}^\nu+ \Lambda^2\overline{C}_q(t)g^{\mu \nu}\ ,
\label{tmunumatred}
\end{eqnarray}
where $\Lambda$ denotes a nonperturbative mass scale in QCD,
and the matrix element of the gluon part (\ref{totalemt}) is given by the similar parameterization with $q\to g$. 
The dimensionless Lorentz-invariant coefficients, ${\Theta}_{1q}(t), {\Theta}_{2q}(t), \overline{C}_q(t)$, ${\Theta}_{1g}(t), {\Theta}_{2g}(t), \overline{C}_g(t)$, are the gravitational form factors.
Based on parity (P) invariance combined with time-reversal (T) invariance, we can show 
\begin{eqnarray}
\langle p'|T_{q,g}^{\mu \nu}|p\rangle=\langle p'|
T_{q,g}^{\mu \nu}|p\rangle^\ast
%=\langle p|
%\left(T_q^{\mu \nu}\right)^\dagger|p'\rangle
=\langle p|
T_{q,g}^{\mu \nu}|p'\rangle\ ,
\label{ptherm}
\end{eqnarray}
and, therefore, 
the gravitational form factors are real quantities, and
(\ref{tmunumatred}) is the most general form satisfying the symmetry constraints.
% under the interchange $p \leftrightarrow p'$.
We also note that the divergenceless property of (\ref{tmn}), $\partial _\mu T^{\mu\nu}(x)=0$, implies
\begin{eqnarray}
\sum_{q=u,d,s,\ldots}\bar{C}_q(t)+\bar{C}_g(t)=0\ . 
\label{barC}
\end{eqnarray}
An example of the other frequently used notations for the form factors is given as~\cite{Polyakov:2018zvc}
\begin{eqnarray}
 \frac{1}{2}{\Theta}_{2q}(t) = 2A_q (t), \;\;\; \frac{1}{2}{\Theta}_{1q}(t)= -\frac{1}{2}D_q(t)\ ,
\;\;\;\Lambda^2\overline{C}_q(t) = m_h^2 \bar{c}_q(t) \ ,
\label{diffnot}
\end{eqnarray}
and similarly for the gluonic contributions.

\section{Constraints from QCD equations of motion}

We investigate the constraints on the contribution of the quark part, (\ref{tmunumatred}), 
using the QCD equations of motion.
The contraction of the LHS of (\ref{tmunumatred}) with $\Delta_\nu=(p'-p)_\nu$ yields, 
\begin{eqnarray}
\lefteqn{\Delta_\nu\langle p'|T_q^{\mu \nu}|p\rangle
=(p'-p)_\nu \frac{1}{4}\langle p'|\left(-\bar{q} i\overleftarrow{D}^\nu\gamma^{\mu}q -\bar{q} i\overleftarrow{D}^\mu\gamma^{\nu} q
+\bar{q}\gamma^{\mu} i\overrightarrow{D}^\nu q+\bar{q}\gamma^{\nu} i\overrightarrow{D}^\mu q\right)|p\rangle}
\nonumber\\
&&\;\;\;\;\;\;\;\;\;\;\;\;\;\;\;\;\;= \frac{1}{4}\langle p'|\left[ \hat{P}_\nu, -\bar{q} i\overleftarrow{D}^\nu\gamma^{\mu}q -\bar{q} i\overleftarrow{D}^\mu\gamma^{\nu} q
+\bar{q}\gamma^{\mu} i\overrightarrow{D}^\nu q+\bar{q}\gamma^{\nu} i\overrightarrow{D}^\mu q \right]|p\rangle
\nonumber\\
&&\;\;\;\;\;\;\;\;\;\;\;\;\;\;\;\;\;=-\frac{1}{4}\langle p'|i \partial_\nu\left\{  -\bar{q} i\overleftarrow{D}^\nu\gamma^{\mu}q -\bar{q} i\overleftarrow{D}^\mu\gamma^{\nu} q
+\bar{q}\gamma^{\mu} i\overrightarrow{D}^\nu q+\bar{q}\gamma^{\nu} i\overrightarrow{D}^\mu q\right\}|p\rangle
\ ,
\label{deltatmunu}
\end{eqnarray}
where 
%we note, e.g., 
%\begin{eqnarray}
%(p'-p)^\mu \langle p'|\bar{q}\gamma^{\mu} i\overrightarrow{D}^\nu q|p\rangle=\langle p'|\left[ \hat{P}^\mu, \bar{q}\gamma^{\mu} %i\overrightarrow{D}^\nu q\right]|p\rangle=-\langle p'|i \partial^\mu\left( \bar{q}\gamma^{\mu} i\overrightarrow{D}^\nu q\right)|%p\rangle\ ,
%\label{mv2}
%\end{eqnarray}
we used the Heisenberg equations for the field operators
with the 4-momentum operator $\hat{P}^\mu$ of QCD, with
\begin{eqnarray}
\hat{P}^\mu = \int d^3x T^{\mu 0}(x)\ , \;\;\;\;\;\hat{P}^\mu|p\rangle=p^\mu |p\rangle\ ,
\end{eqnarray}
and we introduced a shorthand notation
for the derivative over the total translation as
\begin{equation}
\partial_{\alpha}\left\{-\bar{q}(x) i\overleftarrow{D}^\nu\gamma^{\mu}q(x)\right\} \equiv
\left. \frac{\partial}{\partial y^{\alpha}}
\left\{ -\bar{q}(x+y) i\overleftarrow{D}^\nu\gamma^{\mu}q(x+y) \right\} \right|_{y \rightarrow 0}\ ,
\label{eq:3tdrv}
\end{equation}
and so on,
and the corresponding total derivative is evaluated as
\begin{eqnarray}
\lefteqn{-i\partial_\nu \left\{-\bar{q} i\overleftarrow{D}^\nu\gamma^{\mu}q\right\}
=\bar{q} i\overleftarrow{D}^\nu \left( i\overleftarrow{\partial}_\nu + i\overrightarrow{\partial}_\nu \right)\gamma^{\mu}q}
%= -i\partial_\nu \left(-\bar{q} i\overleftarrow{D}^\nu\gamma^{\mu}q\right)
%\nonumber\\
%&=&\bar{q} i\overleftarrow{D}^\nu \left( i\overleftarrow{D}_\nu + i\overrightarrow{D}_\nu \right)\gamma^{\mu}q
%\nonumber\\
%&=&-\bar{q} \overleftarrow{D}^\nu \overleftarrow{D}_\nu \gamma^{\mu}q
%+ \bar{q} i\overleftarrow{D}^\nu i\overrightarrow{D}_\nu \gamma^{\mu}q
%\nonumber\\
%&=&-\bar{q} \overleftarrow{D}_\nu \overleftarrow{D}_\alpha g^{\nu \alpha}\gamma^{\mu}q
%+ \bar{q} i\overleftarrow{D}^\nu i\overrightarrow{D}_\nu \gamma^{\mu}q
\nonumber\\
&&\;\;\;\;\;\;\;\;\;\;\;\;\;\;\;\;\;\;\;=-\bar{q} \overleftarrow{D}_\nu \overleftarrow{D}_\alpha \left(\gamma^\nu \gamma^\alpha+i\sigma^{\nu \alpha}\right)\gamma^{\mu}q
+ \bar{q} i\overleftarrow{D}^\nu i\overrightarrow{D}_\nu \gamma^{\mu}q
%\nonumber\\
%&=&\bar{q} i\overleftarrow{\Slash{D}} i\overleftarrow{\Slash{D}} \gamma^{\mu}q
%-\bar{q} \overleftarrow{D}_\nu \overleftarrow{D}_\alpha i\sigma^{\nu \alpha}\gamma^{\mu}q
%+ \bar{q} i\overleftarrow{D}^\nu i\overrightarrow{D}_\nu \gamma^{\mu}q
%\nonumber\\
%&=&\bar{q} \left(i\overleftarrow{\Slash{D}}+m-m\right) i\overleftarrow{\Slash{D}} \gamma^{\mu}q
%-\bar{q} \left(\frac{1}{2}\left[\overleftarrow{D}_\nu , \overleftarrow{D}_\alpha \right]+\frac{1}{2}\left\{\overleftarrow{D}_\nu , %\overleftarrow{D}_\alpha \right\}\right)i\sigma^{\nu \alpha}\gamma^{\mu}q
%\nonumber\\
%&&+ \bar{q} i\overleftarrow{D}^\nu i\overrightarrow{D}_\nu \gamma^{\mu}q
\nonumber\\
&&\;\;\;\;\;\;\;\;\;\;\;\;\;\;\;\;\;\;\;=-m\bar{q} \left(i\overleftarrow{\Slash{D}}+m-m\right) \gamma^{\mu}q
-\frac{1}{2}\bar{q} \left[\overleftarrow{D}_\nu , \overleftarrow{D}_\alpha \right]i\sigma^{\nu \alpha}\gamma^{\mu}q
+ \bar{q} i\overleftarrow{D}^\nu i\overrightarrow{D}_\nu \gamma^{\mu}q+{\rm EOM}
%\nonumber\\
%&=&m^2\bar{q}  \gamma^{\mu}q
%-\frac{1}{2}\bar{q}  \left(-ig F_{\nu \alpha}\right)i\sigma^{\nu \alpha}\gamma^{\mu}q
%+ \bar{q} i\overleftarrow{D}^\nu i\overrightarrow{D}_\nu \gamma^{\mu}q+{\rm EOM}
\nonumber\\
&&\;\;\;\;\;\;\;\;\;\;\;\;\;\;\;\;\;\;\;=-\frac{1}{2}\bar{q}  g F_{\nu \alpha}\sigma^{\nu \alpha}\gamma^{\mu}q+ m^2\bar{q}  \gamma^{\mu}q
+ \bar{q} i\overleftarrow{D}^\nu i\overrightarrow{D}_\nu \gamma^{\mu}q+{\rm EOM}
\ ,
\end{eqnarray}
using 
%the identity,  $g^{\nu\alpha}=\gamma^\nu \gamma^\alpha+i\sigma^{\nu \alpha}$,
%from (\ref{gammaid}),
%\begin{eqnarray}
$\left[ \overleftarrow{D}^\mu, \overleftarrow{D}^\nu\right]
=\left[ \overleftarrow{\partial}^\mu +igA^\mu, \overleftarrow{\partial}^\nu +igA^\nu\right]
= -ig F^{\mu \nu}$.
%\end{eqnarray}
Here, ``EOM'' denotes the operators that vanish by the use of the equations of motion,
$\bar{q} \left(i\overleftarrow{\Slash{D}}+m\right)=0$.
Combining this with the other terms of (\ref{deltatmunu}) evaluated similarly, 
%the second, third, and forth terms are evaluated as,
we eventually obtain the significantly compact formula~\cite{Kolesnichenko:1984dj,Braun:2004vf},
\begin{eqnarray}
\Delta_\nu\langle p'|T_q^{\mu \nu}|p\rangle
=\langle p'|\bar{q}ig F^{\mu \nu}\gamma_\nu q|p\rangle
\ ,
\label{rr1}
\end{eqnarray}
and, using (\ref{tmunumatred}), we find
\begin{eqnarray}
 \Delta^\mu \Lambda^2\overline{C}_q(t)=\langle p'|\bar{q}ig F^{\mu \nu}\gamma_\nu q|p\rangle\ ,
%\;\;\;\;\left(\mbox{leading to,     } \;\;\; =f_3(t) \right)\ ,
\label{f3theta0q}
\end{eqnarray} 
which shows that $\Lambda^2\overline{C}_q(t)$ is related to quark-gluon interactions corresponding to twist four and higher.
Combined with (\ref{barC}), this also allows us to obtain,
\begin{eqnarray}
\Delta^\mu \Lambda^2\overline{C}_g(t) &=&-\sum_q\Delta^\mu\Lambda^2\overline{C}_q(t)= -\sum_q\langle p'|\bar{q}ig F^{\mu \nu}\gamma_\nu q|p\rangle
\nonumber\\
&&= \langle p'|i F_a^{\mu \nu}\left(-\sum_qg\bar{q}t^a\gamma_\nu  q\right)|p\rangle
=\langle p'| F_a^{\mu \nu} iD^\rho_{ab}  F^b_{\rho \nu}|p\rangle\ ,
\label{f3theta0g}
\end{eqnarray}
using the QCD EOM for the gluon,
$\left[D_\mu ,  F^{\mu \nu}\right]=-t^a\sum_{q'} g\bar{q}'t^a \gamma^\nu q'$.

We may contract this with $\Delta_\mu$ further and perform the manipulations similarly as above,
using the QCD EOM.
This yields
\begin{eqnarray}
%\color{red}
t\Lambda^2\overline{C}_q(t)
&=&
\Delta_\mu\langle p'|\bar{q}ig F^{\mu \nu}\gamma_\nu q|p\rangle
=\langle p'|\partial_\mu\left(\bar{q}g F^{\mu \nu}\gamma_\nu q\right)|p\rangle
\nonumber\\
&&\!\!\!\!\!\!\!\!\!\!\!\!\!\!\!\!\!\!=
- \langle p'|g^2\bar{q}t^a\gamma_\nu q \sum_{q'} \bar{q}'t^a \gamma^\nu q'
|p\rangle
+
 \langle p'|\left(\bar{q}\overleftarrow{D}_\mu g F^{\mu \nu}\gamma_\nu q+ \bar{q}g F^{\mu \nu}\gamma_\nu \overrightarrow{D}_\mu q\right)
|p\rangle
\ ,
\label{theta0qfmunu}
\end{eqnarray}
which represents $t \times \Lambda^2\overline{C}_q(t)$ as matrix elements of the four-quark correlation and quark-gluon correlation effects
of twist six.
Note that the PT invariance implies
\begin{eqnarray}
\langle p'|\bar{q}g F^{\mu \nu}\gamma_\nu \overrightarrow{D}_\mu q|p\rangle
=\langle p'|\bar{q}g F^{\mu \nu}  \gamma_\nu
\overrightarrow{D}_\mu q |p\rangle^\ast
%=\langle p|\left(\bar{q}g F^{\mu \nu}  \gamma_\nu
%\overrightarrow{D}_\mu q \right)^\dagger|p'\rangle
=\langle p| \bar{q}\overleftarrow{D}_\mu g F^{\mu \nu}  \gamma_\nu
 q |p'\rangle
\ ,
\label{0ptpt}
\end{eqnarray}
%which shows that the matrix element $\langle p'|T_q^{\mu \nu}|p\rangle$ is symmetric under the
%interchange $p \leftrightarrow p'$;
therefore, the last two terms in (\ref{theta0qfmunu}) are unlikely to cancel or vanish.

On the other hand, contracting both sides of (\ref{tmunumatred}) with the metric tensor $g_{\mu \nu}$ yields
\begin{eqnarray}
&& \frac{1}{4}\langle p'|\left(-\bar{q} i\overleftarrow{D}\cdot \gamma q -\bar{q} i\overleftarrow{D}\cdot \gamma q
+\bar{q}\gamma\cdot  i\overrightarrow{D} q+\bar{q}\gamma\cdot i\overrightarrow{D} q\right)|p\rangle
\nonumber\\
&&=\frac{1}{2}{\Theta}_{1q}(t) \times 3t +\frac{1}{2} {\Theta}_{2q}(t) \overline{P}^2
 +4 \Lambda^2\overline{C}_q(t)\ ,
\end{eqnarray}
and, using the QCD equations of motion, 
$\bar{q}\left( i\overleftarrow{\Slash{D}} +m\right)=\left(i\overrightarrow{\Slash{D}}-m\right)q=0$,  we obtain,
\begin{eqnarray}
\langle p'| 
m\bar{q} q|p\rangle
= \frac{3t}{2}{\Theta}_{1q}(t)+ 4 \Lambda^2\overline{C}_q(t)+ \frac{1}{2}{\Theta}_{2q}(t) \left( m_h^2 -\frac{t}{4} \right)\ ,
\end{eqnarray}
but the LHS corresponds to the trace of the energy-momentum tensor and its flavor singlet part is modified by
the trace anomaly~\cite{Collins:1976yq,Nielsen:1977sy}, which eventually leads to the relation,
\begin{eqnarray}
\lefteqn{\frac{3t}{2}\left(\sum_q   {\Theta}_{1q}(t) + {\Theta}_{1g}(t)\right)
+\frac{1}{2}\left( m_h^2 -\frac{t}{4} \right)\left( \sum_q {\Theta}_{2q}(t)+{\Theta}_{2g}(t)\right)}
\nonumber\\
&&=\sum_q\left(1+\gamma_m\right) \langle p'| 
m\bar{q} q|p\rangle+
\langle p'|  \frac{\beta(g)}{2g}F_a^{\mu\nu}{F_a}_{\mu\nu}|p\rangle
\ ,
%\nonumber\\
%&&\sum_q m_q\bar{q} q \to\left(1+\gamma_m\right) \sum_q  m\bar{q} q\;\; \; \left(+\;\; \; \frac{\beta(g)}{2g}F_a^{\mu\nu}%{F_a}_{\mu\nu}\right)\ ,
\label{theta1q0qgsummary}
\end{eqnarray}
using  (\ref{barC}), where $\gamma_m$ is the anomalous dimension of the mass operator, 
and $\beta(g)$ is the $\beta$ function of QCD.
This relation for the sum over all partons could serve as a consistency check of the results. For example, taking the $t\to 0$ limit of this relation
and using the sum rule in the forward limit,
\begin{eqnarray}
\sum_q {\Theta}_{2q}(0)+{\Theta}_{2g}(0)=4\ ,
\label{theta2qsr}
\end{eqnarray}
due to the fact that 
%\begin{equation}
$\langle p|T^{\mu\nu}|p\rangle =2p^\mu p^\nu$ holds for (\ref{tmn}),
%\end{equation}
representing the total energy-momentum~\cite{Jaffe:1989jz},
we get,
\begin{eqnarray}
\!\!\!\!\!\!\!\!\!\!\!
\frac{1}{2}\left( \sum_q {\Theta}_{2q}(0)+{\Theta}_{2g}(0)\right) m_h^2  =2m_h^2
=\left\langle p\left| \left(\sum_q\left(1+\gamma_m\right) 
m\bar{q} q+
 \frac{\beta(g)}{2g}F_a^{\mu\nu}{F_a}_{\mu\nu}
\right)\right|p\right\rangle\ ,
%\nonumber\\
%&&\sum_q m_q\bar{q} q \to\left(1+\gamma_m\right) \sum_q  m\bar{q} q\;\; \; \left(+\;\; \; \frac{\beta(g)}{2g}F_a^{\mu\nu}%{F_a}_{\mu\nu}\right)\ ,
\label{theta1q0qgsummarylimit}
\end{eqnarray}
which reproduces the well-known result for the mass composition (see, e.g., \cite{Lorce:2017xzd}); therefore, (\ref{theta1q0qgsummary}) 
corresponds to an off-forward 
generalization of (\ref{theta1q0qgsummarylimit}).

Before ending this section, we note an important exact operator identity, 
\begin{eqnarray}
\frac{1}{2}  
\bar{q}(0) \gamma^{[\mu}\overleftrightarrow{D}^{\alpha ]}  q(0)
= \frac{1}{4}\epsilon^{\mu \alpha \rho \nu} i\partial_\rho\left\{\bar{q}(0)\gamma_\nu\gamma_5 q(0)\right\}
 +{\rm EOM}\ ,
\label{exactlocalid00}
\end{eqnarray}
and its consequence on matrix elements
relevant for the investigation of the gravitational form factor. 
Here, $[\mu \alpha]$ implies the antisymmetrization for the indices $\mu, \alpha$ as
\begin{eqnarray}
\frac{1}{2}  
\bar{q} \gamma^{[\mu}\overleftrightarrow{D}^{\alpha ]} q
=\frac{1}{4}\bar{q} \left(\gamma^{\mu}\overleftrightarrow{D}^{\alpha}-  \gamma^{\alpha}\overleftrightarrow{D}^{\mu} \right)  q\ .
\end{eqnarray}
Actually, this identity has been mentioned repeatedly in many previous papers
and can be derived directly performing the relevant gamma matrix algebra: 
\begin{eqnarray}
\frac{1}{2}  
\bar{q} \gamma^{[\mu}\overleftrightarrow{D}^{\alpha ]}  q
%&=&\frac{1}{8}\bar{q} \left(\gamma^{\mu}\left\{ \gamma^\alpha ,\overleftrightarrow{\Slash{D}}\right\}- \left\{ %\gamma^\mu ,\overleftrightarrow{\Slash{D}}\right\} \gamma^{\alpha} \right)  \Gamma_5 q
%\nonumber\\
%&=&\frac{1}{8}\bar{q} \left(\gamma^{\mu}\left(\gamma^\alpha \overleftrightarrow{\Slash{D}}
%+\overleftrightarrow{\Slash{D}}\gamma^\alpha \right)- 
%\left( \gamma^\mu \overleftrightarrow{\Slash{D}}+\overleftrightarrow{\Slash{D}}\gamma^\mu \right) \gamma^{\alpha} %\right)   q
%\nonumber\\
&=&\frac{1}{8}\bar{q} \left(\gamma^{\mu}\gamma^\alpha \overleftrightarrow{\Slash{D}}
- 
\overleftrightarrow{\Slash{D}}\gamma^\mu  \gamma^{\alpha} \right)  q
\nonumber\\
&=&\frac{i}{8}\bar{q} \left(\sigma^{\mu \alpha}\overleftarrow{\Slash{D}}
+
\overrightarrow{\Slash{D}}\sigma^{\mu \alpha}\right) q
-\frac{i}{8}\bar{q} \left(\sigma^{\mu \alpha}\overrightarrow{\Slash{D}}
+\overleftarrow{\Slash{D}}\sigma^{\mu \alpha}\right)  q
\ ,
\end{eqnarray}
and, noting that the last term equals the EOM operators plus the quark mass term,
we get,
\begin{eqnarray}
\frac{1}{2}  
\bar{q} \gamma^{[\mu}\overleftrightarrow{D}^{\alpha ]}  q
&=&\frac{i}{8}\bar{q} \left(\sigma^{\mu \alpha}\overleftarrow{\Slash{D}}
+
\overrightarrow{\Slash{D}}\sigma^{\mu \alpha}\right)   q
+\frac{i}{8}\bar{q} \left(\sigma^{\mu \alpha}\overrightarrow{\Slash{D}}
+\overleftarrow{\Slash{D}}\sigma^{\mu \alpha}\right)   q
%\nonumber\\
%&&
-2\times\frac{i}{8}\bar{q} \left(\sigma^{\mu \alpha}\overrightarrow{\Slash{D}}
+\overleftarrow{\Slash{D}}\sigma^{\mu \alpha}\right)  q
\nonumber\\
&=&\frac{i}{8}\partial_\nu \left\{\bar{q} \left(\sigma^{\mu \alpha}\gamma^\nu+
\gamma^\nu\sigma^{\mu \alpha}
\right)   q\right\}
-\frac{i}{4}\bar{q} \left(\sigma^{\mu \alpha}\overrightarrow{\Slash{D}}
+\overleftarrow{\Slash{D}}\sigma^{\mu \alpha}\right)  q
\ ,
\end{eqnarray}
whose last two terms cancel out using the EOM, even when including the quark mass effect,
and thus we obtain the exact identity (\ref{exactlocalid00}).\footnote{
Alternatively, this can be derived by Taylor expanding the corresponding nonlocal operator identity 
obtained in~\cite{Balitsky:1987bk}.}
As a consequence of this identity, we find
\begin{eqnarray}
\left\langle p' \left| \frac{1}{2} \bar{q} \gamma^{[\mu} \overleftrightarrow{D}^{\alpha]} q
 \right| p \right\rangle
=
\frac{1}{4}\epsilon^{\mu \alpha \rho \nu}(-\Delta_\rho)\left\langle p' \left|  \bar{q}  \gamma_\nu\gamma_5 q
 \right| p \right\rangle
=0\ ,
\label{identityconseq}
\end{eqnarray}
because the matrix element of the local axial-vector current vanishes for a spin-0 hadron.

\section{Twist-three and twist-four GPDs for spin-0 hadron}

The twist-two GPD for a spin-0 hadron $h$ is defined as~\cite{Polyakov:1998ze,Polyakov:1999gs}
\begin{eqnarray}
\lefteqn{\int_{-\infty}^\infty  \frac{d y^-}{4\pi}e^{i x \overline{P}^+ y^-}
 \left\langle p' \left| 
 \bar{q}(-y/2) \gamma^+ q(y/2) 
 \right| p \right\rangle \Big |_{y^+ = \bm{y}_\perp =0}}
\nonumber\\
&&=
 \int_{-\infty}^\infty  \frac{d \lambda}{4\pi}e^{i x \lambda}
 \left\langle p' \left| 
 \bar{q}(-\lambda n/2) \Slash{n} q(\lambda n/2) 
 \right| p \right\rangle 
=  
H^q_{2} (x,\eta,t)\ ,
\label{pigpd}
\end{eqnarray}
for each separate quark flavor.
Here, $x$ denotes the average longitudinal momentum fraction, and
$H^q_{2}(x,\eta,t)$ is usually denoted simply as $H^q(x,\eta,t)$.
For convenience, we have introduced  lightlike vectors $n^\mu=g^\mu_-n^-$
and $\widetilde{n}^\mu=g^\mu_+\widetilde{n}^+$,
%we employed the light-like vectors $n^\mu$ and $\widetilde{n}^\mu$ 
satisfying
\begin{eqnarray}
n^2=\widetilde{n}^2=0\ ,\;\;\; n\cdot \widetilde{n}=1\ ,
\label{kinema31}
\end{eqnarray}
such that $y^-=\lambda n^-$ with $\overline{P}^+n^-=1$,
and skewness is defined as
\begin{eqnarray}
%t=\Delta^2\ ,\;\;\; 
%\xi=\frac{-\overline{q}^2}{2\overline{P}\cdot \overline{q}}\ ,\;\;\; 
\eta=\frac{-\Delta \cdot n}{2 \overline{P}\cdot n}
%=\frac{-\Delta \cdot \overline{q}}{2 \overline{P}\cdot\overline{q}}
%+{\cal O}(\mbox{twist-}4)
\ ,
\label{kinema20}
\end{eqnarray}
and we have the decompositions of the relevant momenta into the light cone and the perpendicular components as
\begin{eqnarray}
\overline{P}_\mu&=&\widetilde{n}_\mu+\frac{1}{2}\left(m_h^2-\frac{t}{4}\right)n_\mu
\ ,
\nonumber\\
\Delta_\mu &=&-2\eta \widetilde{n}_\mu+\eta\left(m_h^2-\frac{t}{4}\right) n_\mu +\Delta_{\perp \mu}\ ,
\label{kinema40}
\end{eqnarray}
yielding
\begin{eqnarray}
\Delta_\perp^2=-\bm{\Delta}_\perp^2
%=t+4\left(m_h^2-\frac{t}{4}\right)\eta^2
=(1-\eta^2)t +4\eta^2m_h^2\ . 
\label{kinema50}
\end{eqnarray}
Because we now have three independent 4-vectors,
$\widetilde{n}_\mu$, $\Delta_{\perp\mu}$, and $n_\mu$,
%with $\Delta_\perp^\mu \equiv \Delta^\mu -(\Delta\cdot \tilde{n}) n^\mu-(\Delta\cdot n) \tilde{n}^\mu$
%(see (\ref{kinema40})), to parametrize the matrix element in the LHS, 
(\ref{pigpd}) may be generalized as
\begin{eqnarray}
 \lefteqn{\int_{-\infty}^\infty  \frac{d \lambda}{4\pi}e^{i x \lambda}
 \left\langle p' \left| 
 \bar{q}(-\lambda n/2) \gamma^\mu q(\lambda n/2) 
 \right| p \right\rangle}
\nonumber\\
&&=  
\tilde{n}^\mu H^q_{2} (x,\eta,t)+  \Delta_\perp^\mu H^q_{3}(x,\eta,t) + n^\mu H_{4}^q (x,\eta,t)\ ,
\label{defh234pre}
\end{eqnarray}
i.e.,
\begin{eqnarray}
 \lefteqn{
\left\langle p' \left| 
 \bar{q}(-\lambda n/2) \left[ -\frac{\lambda n}{2}, \frac{\lambda n}{2}\right]\gamma^\mu q(\lambda n/2) 
 \right| p \right\rangle }
%\nonumber\\
%&=&  
%2\left(\tilde{n}^\mu \int dx e^{-ix\lambda} H^q_{2} (x,\eta,t)+  \Delta_\perp^\mu \int dx e^{-ix\lambda} H^q_{3}(x,\eta,t) + %n^\mu \int dx e^{-ix\lambda} H_{4}^q (x,\eta,t)
%\right)
\nonumber\\
&&\!\!\!\!\!\!\!\!\!\!\!
=  
2\tilde{n}^\mu \int_{-1}^1 dx e^{-ix\lambda} H^q_{2} (x,\eta,t)+ 2 \Delta_\perp^\mu \int_{-1}^1 dx e^{-ix\lambda} H^q_{3}(x,\eta,t) 
+ 2n^\mu \int_{-1}^1 dx e^{-ix\lambda} H_{4}^q (x,\eta,t)
\ ,
\label{defh234}
\end{eqnarray}
with the path-ordered gauge factor along the straight line 
connecting points $x$ and $y$,
\begin{equation}
[x,y] =\mbox{\rm Pexp}[ig\!\!\int_0^1\!\! dt\,(x-y)_\mu A^\mu(tx+(1-t)y)]\ ,
\label{Pexp}
\end{equation}
being shown explicitly in between the bilocal quark fields on the LHS.
Here, the formal counting of twist for $H^q_{3}(x,\eta,t)$ and $H_{4}^q (x,\eta,t)$ are twist three and four,
respectively. This definition for the twist-three GPD $H^q_{3}(x,\eta,t)$ corresponds to Eq.~(5.69) in \cite{Belitsky:2005qn}
or Eq.(13) in \cite{Anikin:2000em} (see also \cite{Belitsky:2000vx,Kivel:2000rb}).
% {\it {\color{red} [This should be checked!]}}.
It is straightforward to see that these GPDs are constrained from the PT invariance,
implying
\begin{eqnarray}
%\color{red}
\left\langle p' \left| 
 \bar{q}(-\lambda n/2) \left[ -\frac{\lambda n}{2}, \frac{\lambda n}{2}\right]\gamma^\mu q(\lambda n/2) 
 \right| p \right\rangle
&=&\left\langle p' \left|\bar{q}(\lambda n/2) \left[ \frac{\lambda n}{2}, -\frac{\lambda n}{2}\right]
\gamma^\mu q(-\lambda n/2) \right| p \right\rangle^\ast
\nonumber\\
&&\!\!\!\!\!\!\!\!\!\!\!\!\!\!\!\!\!\!\!\!
= \left\langle p\left| 
 \bar{q}(-\lambda n/2) \left[ -\frac{\lambda n}{2}, \frac{\lambda n}{2}\right]\gamma^\mu q(\lambda n/2) 
 \right| p' \right\rangle\ ,
\label{ptbl}
\end{eqnarray}
which shows 
that $ H^q_{2} (x,\eta,t)$, $ H^q_{3} (x,\eta,t)$, and $ H^q_{4} (x,\eta,t)$ are real functions, 
satisfying the symmetry properties as
\begin{eqnarray}
H^q_{2} (x,\eta,t)&=&H^q_{2} (x,-\eta,t)\ ,\;  \;\;\;\;\;\;\;
H^q_{4} (x,\eta,t)=H^q_{4} (x,-\eta,t)\ ,
\nonumber\\
H^q_{3} (x,\eta,t)&=&
-H^q_{3} (x,-\eta,t)\ .
\label{oddeven}
\end{eqnarray}
Now,
Taylor expanding both sides of (\ref{defh234}) about $\lambda= 0$, we obtain
\begin{eqnarray}
 \lefteqn{
 \left\langle p' \left| 
 \bar{q}(0) \gamma^\mu q(0) 
 \right| p \right\rangle+\frac{\lambda}{2} n_\nu\left\langle p' \left| 
 \bar{q}(0) \gamma^\mu \overleftrightarrow{D}^\nu q(0) 
 \right| p \right\rangle+\cdots} 
\nonumber\\
&=&  
2\tilde{n}^\mu \int_{-1}^1 dx  H^q_{2} (x,\eta,t)+ 2 \Delta_\perp^\mu \int_{-1}^1 dx  H^q_{3}(x,\eta,t) 
+ 2n^\mu \int_{-1}^1 dx  H_{4}^q (x,\eta,t)
\nonumber\\
&&\!\!\!\!\!\!\!\!\!\!\!\!\!\!\!\!\!\!\!
-i\lambda\left[2\tilde{n}^\mu \int_{-1}^1 dx x H^q_{2} (x,\eta,t)+ 2 \Delta_\perp^\mu \int_{-1}^1 dx  xH^q_{3}(x,\eta,t) 
+ 2n^\mu \int_{-1}^1 dx x H_{4}^q (x,\eta,t)\right]
+ \cdots\ ,
\end{eqnarray}
so that
\begin{eqnarray}
\left\langle p' \left| 
 \bar{q}(0) \gamma^\mu q(0) 
 \right| p \right\rangle&=&2\tilde{n}^\mu \int_{-1}^1 dx  H^q_{2} (x,\eta,t)+ 2 \Delta_\perp^\mu \int_{-1}^1 dx  H^q_{3}(x,\eta,t) 
 \nonumber\\
&+& 2n^\mu \int_{-1}^1 dx  H_{4}^q (x,\eta,t)\ ,
\label{pigpdmom0}\\
 n_\nu\left\langle p' \left| \frac{1}{2}
 \bar{q}(0) \gamma^\mu i \overleftrightarrow{D}^\nu q(0) 
 \right| p \right\rangle&=&2\tilde{n}^\mu \int_{-1}^1 dx x H^q_{2} (x,\eta,t)+ 2 \Delta_\perp^\mu \int_{-1}^1 dx  xH^q_{3}(x,\eta,t) 
 \nonumber\\
&&
%\;\;\;\;\;\;\;\;\;\;\;\;\;\;\;\;\;\;\;\;\;\;\;\;\;\;\;\;\;\;\;\;\;\;\;\;\;\;\;\;\;\;\;\;\;
+ 2n^\mu \int_{-1}^1 dx x H_{4}^q (x,\eta,t)\ ,
\label{pigpdmom1}
\end{eqnarray}
and similarly for higher moments. 
Denoting the quark contribution to the vector form factor of the spin-0 hadron 
as ${\cal F}_v(\Delta^2)$ and substituting (\ref{kinema40}), the LHS of 
(\ref{pigpdmom0}) is expressed as, 
\begin{eqnarray}
\langle p'|\bar{q}  \gamma_\mu q|p\rangle&=&2 {\cal F}_v(t)\overline{P}_\mu=
2{\cal F}_v(t)\left[\widetilde{n}_\mu+\frac{1}{2}\left(m_h^2-\frac{t}{4}\right)n_\mu\right]
\ ; 
\label{vform}
\end{eqnarray}
therefore, we find
\begin{eqnarray}
&&2 \int_{-1}^1 dx  H^q_{2} (x,\eta,t)=2{\cal F}_v(t)\ ,\;\;\; \int_{-1}^1 dx  H^q_{3}(x,\eta,t)=0\ ,
\nonumber\\
&& 2 \int_{-1}^1 dx  H_{4}^q (x,\eta,t)=\left(m_h^2-\frac{t}{4}\right){\cal F}_v(t)\ .
\label{calmv}
\end{eqnarray}
For the operator in the LHS of (\ref{pigpdmom1}), we note, using (\ref{symtmnq}), that
\begin{eqnarray}
\frac{1}{2} \bar{q}(0) \gamma^\mu i \overleftrightarrow{D}^\nu q(0) &=&
\frac{1}{2} \left( \bar{q}(0) \gamma^{(\mu} i \overleftrightarrow{D}^{\nu)} q(0) +  \bar{q}(0) \gamma^{[\mu} i \overleftrightarrow{D}^{\nu]} q(0)\right)
\nonumber\\
&=&T_q^{\mu \nu}(0)+ \frac{1}{2} \bar{q}(0) \gamma^{[\mu} i \overleftrightarrow{D}^{\nu]} q(0)\ ,
\end{eqnarray}
and, combined with (\ref{identityconseq}), we obtain
\begin{eqnarray}
 n_\nu\left\langle p' \left| T_q^{\mu \nu}(0)
 \right| p \right\rangle
%+  n_\nu\left\langle p' \left| \frac{1}{2} \bar{q}(0) \gamma^{[\mu} i \overleftrightarrow{D}^{\nu]} q(0)
% \right| p \right\rangle
% \nonumber\\
 &&=2\tilde{n}^\mu \int_{-1}^1 dx x H^q_{2} (x,\eta,t)+ 2 \Delta_\perp^\mu \int_{-1}^1 dx  xH^q_{3}(x,\eta,t) 
 \nonumber\\
&&
+ 2n^\mu \int_{-1}^1 dx x H_{4}^q (x,\eta,t)\ .
\label{gpdmom10}
\end{eqnarray}
Substituting the formula (\ref{tmunumatred}) in terms of the gravitational form factors into the 
%first term in the 
LHS and using (\ref{kinema20}), (\ref{kinema40}), we obtain
\begin{eqnarray}
n_\nu\langle p'|T_q^{\mu \nu}(0)|p\rangle
&=&\left(t n^\mu+2\eta \Delta^\mu\right)\frac{1}{2}{\Theta}_{1q}(t)+\overline{P}^\mu \frac{1}{2}{\Theta}_{2q}(t)
 +n^\mu\Lambda^2\overline{C}_q(t) .
\nonumber\\
&=&\widetilde{n}^\mu
\left(\frac{1}{2}{\Theta}_{2q}(t)-2\eta^2 {\Theta}_{1q}(t)
\right)
+\Delta_{\perp}^\mu
\eta {\Theta}_{1q}(t)
\nonumber\\
&&\!\!\!\!\!\!\!\!\!\!\!\!
+n^\mu \left[\left\{\frac{t}{2}+\left(m_h^2-\frac{t}{4}\right)\eta^2\right\}{\Theta}_{1q}(t)
+ \frac{1}{4}\left(m_h^2-\frac{t}{4}\right){\Theta}_{2q}(t)+
 \Lambda^2\overline{C}_q(t)
\right]
\ ,
\label{tmunumatnc}
\end{eqnarray}
which leads to
\begin{eqnarray}
&&\!\!\!\!\!\!2\int_{-1}^1 dx x H^q_{2} (x,\eta,t)=
 \frac{1}{2}{\Theta}_{2q}(t)-2\eta^2 {\Theta}_{1q}(t)\ ,
\nonumber\\
&&\!\!\!\!\!\!2  \int_{-1}^1 dx  xH^q_{3}(x,\eta,t) =
\eta {\Theta}_{1q}(t)
%+\frac{1}{\Delta_\perp^2}
%\Delta_{\perp \mu}n_\nu\left\langle p' \left| \frac{1}{2} \bar{q}(0) \gamma^{[\mu} i \overleftrightarrow{D}^{\nu]} q(0)
% \right| p \right\rangle
\ ,
\nonumber\\
&&\!\!\!\!\!\!2 \int_{-1}^1 dx x H_{4}^q (x,\eta,t)=
\left\{\frac{t}{2}+\left(m_h^2-\frac{t}{4}\right)\eta^2\right\}{\Theta}_{1q}(t)
+ \frac{1}{4}\left(m_h^2-\frac{t}{4}\right){\Theta}_{2q}(t)+
\Lambda^2\overline{C}_q(t)\ .
%\nonumber\\
%&&
% &+&\widetilde{n}_\mu n_\nu\left\langle p' \left| \frac{1}{2} \bar{q}(0) \gamma^{[\mu} i \overleftrightarrow{D}^{\nu]} q(0)
%\right| p \right\rangle
%\ ,
\label{mom1relation}
\end{eqnarray} 
Here, the first formula for $H^q_{2} (x,\eta,t)$ is
the analogue of the following formulas for the usual GPDs $H$ and $E$ for the nucleon ($N$),
which we denote as $H^{(N)q}$ and $E^{(N)q}$, respectively:
\begin{eqnarray}
\int_{-1}^{1} dxx H^{(N)q} (x, \eta,t) &=&A^{(N)}_q (t)+  \eta^2D^{(N)}_q(t)\ ,
\nonumber\\
\int_{-1}^{1} dxxE^{(N)q} (x, \eta,t) &=&B^{(N)}_q(t)- \eta^2D^{(N)}_q(t)\ ,
\end{eqnarray}
where $A^{(N)}_{q} (t), B^{(N)}_{q}(t), D^{(N)}_{q}(t)$ are the gravitational form factors for the nucleon,
defined as matrix elements of the energy-momentum tensor (\ref{symtmnq}) in terms of the nucleon state $|N(p, S)\rangle$
with momentum $p$, mass $m_N$, and spin $S$ (see, e.g., \cite{Ji:1996ek,Hatta:2012jm,Polyakov:2018zvc}):
\begin{eqnarray}
 \langle N(p',S')|T_{q,g}^{\mu\nu}(0)|N(p, S)\rangle &=& \bar{u}(p',S')\Bigl[A^{(N)}_{q,g} (t)\gamma^{(\mu}\overline P^{\nu)}
 +B^{(N)}_{q,g}(t)\frac{\overline P^{(\mu}i\sigma^{\nu)\alpha}\Delta_\alpha}{2m_N} \nonumber \\
 && \quad  + D^{(N)}_{q,g}(t)\frac{\Delta^\mu\Delta^\nu -g^{\mu\nu}\Delta^2}{4m_N} + \bar{c}^{(N)}_{q,g}(t)m_Ng^{\mu\nu}\Bigr] u(p,S)
\ .
\label{para}
\end{eqnarray}
It is worth noting that the moment relations (\ref{calmv}) and (\ref{mom1relation}) have forms
consistent with the symmetry properties
(\ref{oddeven}).
We also note that (\ref{calmv}) and (\ref{mom1relation}) satisfy the relations,
\begin{eqnarray}
\int_{-1}^1dx H_3^q(x,\eta, t)&=&0=-\frac{1}{2}\frac{\partial}{\partial \eta}\int_{-1}^1dx H_2^q(x,\eta, t)\ ,
\nonumber\\
\int_{-1}^1dx x H_3^q(x,\eta, t)&=&-\frac{1}{4}\frac{\partial}{\partial \eta}\int_{-1}^1dx xH_2^q(x,\eta, t)\ ,
\end{eqnarray}
which are pointed out in \cite{Kivel:2000rb} (see also \cite{Belitsky:2005qn}).

It is remarkable that the formulas in (\ref{mom1relation}) provide the exact relations 
between the three independent gravitational form factors arising in (\ref{tmunumatred}) and 
the second moment of the three GPDs which form 
a complete set of GPDs associated with the nonlocal quark-antiquark vector operator.
In particular, the second moment of the twist-three GPD $H_3^q$ completely determines the 
gravitational form factor ${\Theta}_{1q}(t)$.
On the other hand, the last formula in (\ref{mom1relation}),  as well as the last formula in  (\ref{calmv}),
reflects the fact that, in general, parts of the twist-four contributions 
%to the twist-four GPDs 
are given by the lower-twist quantities
multiplied by the kinematic invariants like (\ref{kinema21}) associated with the hadron states,
while the remaining parts correspond to the ``dynamical (genuine)'' twist-four contributions.\footnote{A systematic method to identify and resum the former parts (``kinematic'' contributions) 
for all moments (all Lorentz spins of operators) is developed and is applied to obtain 
all kinematic twist-four corrections to the DVCS, see \cite{Braun:2011zr,Braun:2011dg,Braun:2012bg}.}

In the chiral limit ($m\to 0$, $m_h^2\to 0$), some of the relations at the level of twist four 
%and higher 
are simplified,
and the last formula of (\ref{calmv}) and that of (\ref{mom1relation}) reduce to
\begin{eqnarray}
 2 \int_{-1}^1 dx  H_{4}^q (x,\eta,t)&\to&-\frac{t}{4}{\cal F}_v(t)\ ,
 \nonumber\\
 2 \int_{-1}^1 dx x H_{4}^q (x,\eta,t)&\to&
\frac{t}{4}\left(2-\eta^2\right){\Theta}_{1q}(t)
-\frac{t}{16}{\Theta}_{2q}(t)+
 \Lambda^2\overline{C}_q(t)=\Lambda^2\overline{C}_q(0)+{\cal O}(t)\ ,
\label{pointout}
\end{eqnarray}
respectively, where, in the second formula we also display the leading term in the forward limit,
using the fact that ${\Theta}_{1q}(t=0)$ should be finite.
It is also worth mentioning a particular property relevant for pions~\cite{Polyakov:1998ze}:
applying the soft pion theorem for $t\to 0$
allows us to obtain
\begin{equation}
{\Theta}_{1q}(t)=P^q+ {\cal O}(t) + {\cal O}(m_h^2)\ ,
\label{climit}
\end{equation}
with 
\begin{equation}
P^q=\frac{1}{4}{\Theta}_{2q}(0)\ ,
\label{pqtheta2q}
\end{equation}
being the average momentum fraction carried by the quark of 
flavor $q$ (see (\ref{theta2qsr}))
and the correction (${\cal O}(t)$, ${\cal O}(m_h^2)$) terms receive the chiral logarithms~\cite{Donoghue:1991qv,Kubis:1999db}. Comparing this result with (\ref{mom1relation}), 
one obtains~\cite{Kivel:2000rb} (see also \cite{Belitsky:2005qn}).
\begin{equation}
\int_{-1}^1dx x H_3^q(x,\eta, t)=\frac{1}{2} \eta P^q+ {\cal O}(t) + {\cal O}(m_h^2)\ .
\end{equation}

\section{Unravelling in the light-cone gauge fixing}

Exact gauge-invariant manipulations discussed above, utilizing symmetries, EOM, and the relation with the GPDs,
allow us to obtain several constraints on the gravitational form factors in (\ref{tmunumatred}) as well as
the explicit operator content of the form factor $\overline{C}_{q,g}(t)$,
as (\ref{f3theta0q})-(\ref{theta0qfmunu}). However, this approach is not helpful for revealing direct information on 
the form factor ${\Theta}_{1q}(t)$ corresponding to the D term.
To try to assess the operator content of ${\Theta}_{1q}(t)$ particular to the off-forward matrix element,
we employ gauge fixing to allow us to treat each term of the covariant derivative separately and identify the physical degrees 
of freedom.
We take the light-cone gauge, $n_\mu A^\mu=n^- A^+=0$, anticipating manipulations linked with the partonic interpretations 
appropriate in the infinite momentum frame;
in this gauge, 
the gluon field in the covariant derivative can be expressed by the field strength tensor as~\cite{Kogut:1969xa,Ji:1992eu}
\begin{eqnarray}
A^\mu(\lambda n) =\frac{1}{2}\int_{-\infty}^\infty d\lambda'\ {\rm sgn} (\lambda'-\lambda) F^{\mu \alpha}(\lambda' n)n_\alpha\ ,
\label{lcgglue}
\end{eqnarray}
where ${\rm sgn}(\lambda)=\theta(\lambda)-\theta(-\lambda)$,
corresponding to the antiperiodic boundary condition for $A^\mu(y)$ at $|y^-|\to \infty$,
may be replaced with $\frac{1}{2}{\rm sgn} (\lambda) \rightarrow \pm\theta(\pm \lambda)$ for other choices of boundary conditions.
The matrix element of the quark part of the energy-momentum tensor (\ref{symtmnq}) reads, 
\begin{eqnarray}
\langle p'|T_q^{\mu \nu}(y=0)|p\rangle\equiv \langle p'|T_q^{\mu \nu}|p\rangle
=
\frac{1}{4}\langle p'|\bar{q}\left( -i\overleftarrow{\partial}^\mu + i\overrightarrow{\partial}^\mu+2gA^\mu\right)\gamma^{\nu} q
|p\rangle + \left( \mu \leftrightarrow  \nu \right)\ ,
\label{hw0}
\end{eqnarray}
substituting the explicit form of the covariant derivatives.
Here, the derivative terms are handled, using the Heisenberg equations for the quark and antiquark field operators, as
\begin{eqnarray}
\langle p'|\bar{q}\left( -i\overleftarrow{\partial}^\mu + i\overrightarrow{\partial}^\mu\right)\gamma^{\nu} q
|p\rangle
%&=&\langle p'|\left(   \bar{q}\gamma^{\nu}i\overrightarrow{\partial}^\mu q -\bar{q}i\overleftarrow{\partial}^\mu\gamma^{\nu}q \right)
%|p\rangle
%\nonumber\\
&=&\langle p'|\left(   \bar{q}\gamma^{\nu} \left[ q, \hat{\cal P}^\mu\right] -\left[ \bar{q}, \hat{\cal P}^\mu\right]\gamma^{\nu}q \right)
|p\rangle
\nonumber\\
&=&\langle p'|\left(   \bar{q}\gamma^{\nu}  q \hat{\cal P}^\mu+\hat{\cal P}^\mu\bar{q} \gamma^{\nu}q- 2\bar{q} \hat{\cal P}^\mu\gamma^{\nu}q  \right)
|p\rangle
\nonumber\\
&=&2\overline{P}^\mu\langle p'|    \bar{q}\gamma^{\nu}  q|p\rangle
-2\langle p'|\bar{q} \hat{\cal P}^\mu\gamma^{\nu}q|p\rangle\ ,
\label{remove}
\end{eqnarray}
with the 4-momentum operator $\hat{\cal P}^\mu$ in the light-cone quantization of QCD, 
\begin{eqnarray}
\hat{\cal P}^\mu = \int dx^- d^2x_\perp\,  T^{\mu +}(x)\ , \;\;\;\;\;\;\;\;\;\;\hat{\cal P}^\mu|p\rangle=p^\mu |p\rangle\ ,
\end{eqnarray}
and the matrix element in the second term is evaluated by inserting a 
complete set of the light-cone Fock states, $\sum_r|p_r\rangle \langle p_r | =1$, as
\begin{eqnarray}
\langle p'|\bar{q} \hat{\cal P}^\mu\gamma^{\nu}q|p\rangle
=\sum_r p_r^\mu\langle p'|\bar{q} |p_r\rangle \langle p_r | \gamma^{\nu}q|p\rangle
\equiv \langle p_r^\mu \rangle \langle p'|    \bar{q}\gamma^{\nu}  q|p\rangle
\ ,
\label{averagev00}
\end{eqnarray}
where $\langle p_r^\mu \rangle$ denotes the value of $p_r^\mu$ averaged over intermediate states;
this average value would depend in general on $t$.
Because $p_r$ equals $p-k_q$
where $k_q$ denotes the 4-momentum of the
quark removed from the initial state $|p\rangle$ by the action of the field operator $q$,
we may express this matrix element as
\begin{eqnarray}
\langle p'|\bar{q} \hat{\cal P}^\mu\gamma^{\nu}q|p\rangle
\equiv
\langle p^\mu- k_q^\mu\rangle\langle p'|    \bar{q}\gamma^{\nu}  q|p\rangle
=\left(p^\mu-\langle k_q^\mu\rangle\right)\langle p'|    \bar{q}\gamma^{\nu}  q|p\rangle
\ .
\label{averagev}
\end{eqnarray}
Noting that, $k_q^\mu  =x \overline{P}^\mu-\frac{\Delta^\mu}{2}$, in the parton language relevant for the GPD formulation,
which is considered to be appropriate 
for $k_q^\mu$ with $\mu=+$ and $\perp$ when taking (the light-cone quantization in) the light-cone gauge, and assuming that this identification is accurate in the averaging 
for $\langle k_q^\mu \rangle$
in (\ref{averagev}), 
we obtain 
\begin{eqnarray}
%\frac{\sum_n p_n^\mu\langle p'|\bar{q} |p_n\rangle \langle p_n | \gamma^{\nu}q|p\rangle}{\langle p'|\bar{q}\gamma^{\nu}q|%p\rangle}\simeq
p^\mu-\langle  k_q^\mu\rangle\simeq p^\mu- \left\langle x \overline{P}^\mu-\frac{\Delta^\mu}{2}\right\rangle
= \overline{P}^\mu-\langle x\rangle \overline{P}^\mu\ ,
\label{averagex}
\end{eqnarray}
using $p=\overline{P}-\frac{\Delta}{2}$.
Combining this with the above formula (\ref{remove}), we get
\begin{eqnarray}
\langle p'|\bar{q}\left( -i\overleftarrow{\partial}^\mu + i\overrightarrow{\partial}^\mu\right)\gamma^{\nu} q
|p\rangle
\simeq2\langle x\rangle \overline{P}^\mu\langle p'|    \bar{q}\gamma^{\nu}  q|p\rangle
\ ,
\end{eqnarray}
for $\mu=+, \perp$.
Substituting this into (\ref{hw0}), we obtain
\begin{eqnarray}
\langle p'|T_q^{\mu \nu}|p\rangle&\simeq&
\frac{1}{4}\left( 2\langle x\rangle \overline{P}^\mu\langle p'|    \bar{q}\gamma^{\nu}  q|p\rangle+
\langle p'|\bar{q}\ 2gA^\mu\gamma^{\nu} q
|p\rangle\right) + \left( \mu \leftrightarrow  \nu \right)
\nonumber\\
&=&2 \langle x \rangle{\cal F}_v(t) \overline{P}^\mu\overline{P}^\nu
+ \frac{1}{2}\langle p'|\left( \bar{q}gA^\mu\gamma^{\nu}q+ \bar{q}gA^\nu\gamma^{\mu}q\right) 
|p\rangle\ ,
\label{hw}
\end{eqnarray}
where we substituted the definition of the quark contribution of the vector form factor, (\ref{vform}).
Here, $\langle x \rangle$ is the ``average value'' of the quark momentum fraction;
using (\ref{averagev00})
and (\ref{averagex}), it is formally given as
\begin{eqnarray}
\sum_r \frac{p_r^+}{\overline{P}^+}\langle p'|\bar{q} |p_r\rangle \langle p_r | \gamma^{\nu}q|p\rangle
\simeq \left(1-\langle x\rangle \right)\langle p'|\bar{q}\gamma^{\nu}q|p\rangle\ ,
\label{averagedef}
\end{eqnarray}
and the resulting value of $\langle x\rangle$ in general depends on $t$.
Now, comparing (\ref{hw}) with  (\ref{tmunumatred}), we obtain
\begin{eqnarray}
\lefteqn{\frac{1}{2}{\Theta}_{1q}(t)\left(tg^{\mu\nu}- \Delta^\mu \Delta^\nu\right)+ \frac{1}{2}{\Theta}_{2q}(t) \overline{P}^\mu\overline{P}^\nu+\Lambda^2\overline{C}_q(t)g^{\mu \nu}}
\nonumber\\
&\simeq&2\langle x \rangle{\cal F}_v(t) \overline{P}^\mu\overline{P}^\nu
+ \frac{1}{2}\langle p'|\left( \bar{q}gA^\mu\gamma^{\nu}q+ \bar{q}gA^\nu\gamma^{\mu}q\right) 
|p\rangle\ ,
\label{tmunumatredcomp}
\end{eqnarray}
for $\mu=+$ and $\perp$,
with (\ref{lcgglue}) to be substituted for $A^\mu$.
The $\mu=\nu=+$ component of this formula leads to
\begin{eqnarray}
\frac{1}{2}{\Theta}_{2q}(t) -2\eta^2{\Theta}_{1q}(t)
&\simeq&2\langle x \rangle{\cal F}_v(t)\ ,
\label{tmunumatredcompnn}
\end{eqnarray}
and, similarly, taking the component with $\mu=\perp$ and $\nu=+$,
we find
%\begin{eqnarray}
%-{\Theta}_{1q}(t)\Delta^+ \Delta_\perp^\mu&\simeq&
%\langle p'| \bar{q}gA_\perp^\mu\gamma^+ q
%|p\rangle\ , 
%\label{tmunumatredcompdn}
%\end{eqnarray}
%yielding,
\begin{eqnarray}
2 \eta\Delta_\perp^\mu {\Theta}_{1q}(t) &\simeq&
\langle p'| \bar{q}gA_\perp^\mu\Slash{n} q
|p\rangle
\nonumber\\
&=&\frac{1}{2}\int_{-\infty}^\infty d\lambda\ {\rm sgn} (\lambda) 
n_\alpha\langle p'| gF_a^{\mu \alpha}(\lambda n)
\bar{q}(0)t^a\Slash{n} q(0)
|p\rangle
\ .
\label{lctheta1}
\end{eqnarray}
% in particular,
%we note that the following relation holds
%\begin{eqnarray}
%\partial_{\perp\, \mu}\left(\bar{q}g F^{\mu \nu}\Gamma q\right)
%&=&\bar{q}\overleftarrow{D}_{\perp\, \mu} g F^{\mu \nu}\Gamma q
%+\bar{q}\left[D_{\perp\, \mu} , g F^{\mu \nu}\right]\Gamma q
%+\bar{q}g F^{\mu \nu}\Gamma \overrightarrow{D}_{\perp\, \mu} q
%\ ,
%\nonumber\\
%\partial_{\perp\, \mu}\left(\bar{q}g F^{\mu \nu}\Slash{n} q\right)
%&=&\bar{q}\overleftarrow{D}_{\perp\, \mu} g F^{\mu \nu}\Slash{n} q
%+\bar{q}\left[D_{\perp\, \mu} , g F^{\mu \nu}\right]\Slash{n}q
%+\bar{q}g F^{\mu \nu}\Slash{n} \overrightarrow{D}_{\perp\, \mu} q\ ,
%\nonumber\\
%\partial_{\perp\, \mu}n_\alpha\left(\bar{q}g F^{\mu \alpha}\Slash{n} q\right)
%&=&n_\alpha\bar{q}\overleftarrow{D}_{\perp\, \mu} g F^{\mu \alpha}\Slash{n} q
%+n_\alpha\bar{q}\left[D_{\perp\, \mu} , g F^{\mu \alpha}\right]\Slash{n}q
%+n_\alpha\bar{q}g F^{\mu \alpha}\Slash{n} \overrightarrow{D}_{\perp\, \mu} q\ ,
%\end{eqnarray}
%\begin{eqnarray}
%%\color{red}
%2 \eta\Delta_\perp^2 {\Theta}_{1q}(\Delta^2) &\simeq&
%\frac{1}{2}\int d\lambda\ {\rm sgn} (\lambda) 
%\Delta_{\perp\, \mu}n_\alpha\langle p'| \bar{q}gF^{\mu \alpha}(\lambda n)
%\Slash{n} q
%|p\rangle
%\nonumber\\
%&=&
%\ ,
%\label{lctheta11}
%\end{eqnarray}
This expression suggests that, in the parton language appropriate for the infinite momentum frame,
${\Theta}_{1q}(t)$ corresponds to a certain integral of
the twist-three quark-gluon correlation, in particular, 
the correlation between the quark color-current-density and the gluon field strength.
When (\ref{lctheta1}) is summed over the quark flavor $q$,
the result yields
\begin{eqnarray}
\sum_q 2 \eta\Delta_\perp^\mu {\Theta}_{1q}(t)&\simeq&
-\frac{1}{2}\int_{-\infty}^\infty d\lambda\ {\rm sgn} (\lambda) 
n_\alpha\langle p'| F_a^{\mu \alpha}(\lambda n)
n_\nu\sum_q\left(-g\bar{q}(0)t^a\gamma^\nu q(0)\right)
|p\rangle
\nonumber\\
&&=-\frac{1}{2}\int_{-\infty}^\infty d\lambda\ {\rm sgn} (\lambda) 
n_\alpha\langle p'| F_a^{\mu \alpha}(\lambda n)
 D_\rho^{ab}  F_b^{\rho \beta}(0)n_\beta
|p\rangle
\nonumber\\
&=&-\frac{(n^-)^2}{2}\int_{-\infty}^\infty d\lambda\ {\rm sgn} (\lambda) 
\langle p'| F^{\mu +}(\lambda n)
 D_j  F^{j +}(0)
|p\rangle
\nonumber\\
&&-\frac{(n^-)^2}{2}\int_{-\infty}^\infty d\lambda\ {\rm sgn} (\lambda) 
\langle p'| F^{\mu +}(\lambda n)
 D^+  F^{- +}(0)
|p\rangle
\ ,
\label{gluoncorrresult}
\end{eqnarray}
using the gluon EOM, where $j$ is summed over the transverse components, and the final form
is given as an integral of the twist-three bilocal gluon correlation which could be eventually re-expressed 
by the gluonic three-body operators using the technique as in \cite{Kodaira:1997ig,Kodaira:1998jn,Braun:2000yi,Hatta:2012jm}.

The results (\ref{lctheta1}) and (\ref{gluoncorrresult}) originate from the quark-gluon coupling
contained in the covariant derivative of the quark part of the energy-momentum tensor of (\ref{symtmnq}).
In this connection, we note the following point: 
in the canonical formalism, it is not obvious from the outset whether the quark-gluon coupling terms
should be organized as a quark part or not,
because the corresponding terms do not exist in the (nonsymmetric) canonical energy-momentum tensor, but
%the corresponding terms 
those terms are generated as contributions due to the Belinfante improvement 
%of the ``super-potentials'' 
for the gluon part
of the canonical energy-momentum tensor 
(see, e.g., article by Jackiw in \cite{Treiman:1986ep}).
The final forms of the above results,
% of (\ref{lctheta1}) and (\ref{gluoncorrresult}),
corresponding to the current-gluon correlation (\ref{lctheta1}) 
or the gluon-gluon correlation (\ref{gluoncorrresult}), are not surprising in view of this fact.

We note that other components of (\ref{tmunumatredcomp}) with $\mu=
\perp, \nu=\perp$, and $\mu=+, \nu=-$ would give the 
contribution at twist four, but the applicability of 
the partonic relation $k_q = x \overline{P}-\frac{\Delta}{2}$ in the present context would be less obvious
for such components.

The $t=0$ limit of (\ref{tmunumatredcompnn}) reduces to
\begin{eqnarray}
\frac{1}{2}{\Theta}_{2q}(0) -2\eta^2{\Theta}_{1q}(0)
&\simeq&2\left.\langle x \rangle\right|_{t=0}\ ,
\label{tmunumatredcompnn0}
\end{eqnarray}
because ${\cal F}_v(0)=1$, and this result is consistent with the general result (\ref{pqtheta2q}).
On the other hand, the relation of  (\ref{lctheta1}) with the soft-pion result (\ref{climit}) is not obvious
at present,
because it is not straightforward to extract the $t\to 0$ behavior of the QCD matrix element in (\ref{lctheta1}).

If one makes the replacement, $x \to \langle x \rangle$, under the integration in the LHS of the first formula in (\ref{mom1relation})
and combining the result with (\ref{calmv}), 
we obtain the result formally similar to our formula (\ref{tmunumatredcompnn}); it is indeed straightforward to see
that the value $\langle x \rangle$ of (\ref{averagedef}) can be identified as the average value using the $x$ dependence of the corresponding GPD,
based on the operator definition (\ref{defh234}) of the GPDs.
On the other hand, a similar logic is not applicable to the second formula in (\ref{mom1relation}),
because 
(\ref{calmv}) indicates that 
a simple average procedure using the $x$ dependence of $H_3^q(x, \eta, t)$ would not be useful.
This fact suggests the nontrivial nature of our result (\ref{lctheta1}) for ${\Theta}_{1q}(t)$.
Indeed, for the large $t$ behavior of the form factors, our results (\ref{tmunumatredcompnn}),  (\ref{lctheta1}) suggest that
${\Theta}_{2q}(t)$ should obey the dimensional counting rule~\cite{Brodsky:1973kr,Brodsky:1974vy,Matveev:1973ra} which is same as the vector form factor ${\cal F}_v(t)$,
while ${\Theta}_{1q}(t)$ would receive an additional $1/t$ suppression due to the counting for the three-body Fock state.\footnote{The derivation of the counting rule of \cite{Matveev:1973ra} with ``automodellism hypothesis''
suggests that a different counting could be caused only by the different number of the constituents in the participating hadrons.} Namely, because ${\cal F}_v(t\to \infty) \sim 1/t$ as in \cite{Brodsky:1973kr,Brodsky:1974vy,Matveev:1973ra}, we expect the large $t$ behavior as
\begin{eqnarray}
{\Theta}_{2q}(t) \sim \frac{1}{t}\ , \;\;\:\;\;\;\; {\Theta}_{1q}(t) \sim \frac{1}{t^2}\ .
\end{eqnarray}

It is also worth noting somewhat tricky points concerning the twist counting of the form factors:
according to the definition (\ref{tmunumatred}),
${\Theta}_{2q}(t)$ is given by the matrix element of the twist-two operator and 
is thus identified as a twist-two quantity.
Similarly, ${\cal F}_v(t)$ should be a twist-two quantity as matrix element of the 
twist-two operator, see (\ref{vform}).
Combined with (\ref{lctheta1}) which indicates that ${\Theta}_{1q}(t)$ is of
twist three, 
the relation (\ref{tmunumatredcompnn})
implies that ${\Theta}_{2q}(t)$ as well as
${\cal F}_v(t)$ is contaminated by the twist-three contributions.
This mismatch in the twist counting 
is caused by the mismatch between the light-cone ``plus'' direction defined by referring to the average nucleon momentum,
$\overline{P}=\frac{p+p'}{2}$, and the direction collinear to the initial (final) nucleon momentum, 
$p=\overline{P}- \frac{\Delta}{2}$ ($p'=\overline{P} +\frac{\Delta}{2}$); namely,
a light-cone vector along $\tilde n$ has the ``transverse components'' $\propto \Delta_\perp$, when viewed from the initial and final 
nucleons which have the momenta $p$ and $p'$, respectively.

\section{Extensions to the nucleon gravitational form factors}
 
Some of our results discussed above are immediately extended to the corresponding relations for the  
gravitational form factors of the nucleon.
We first note that the exact relations similar to (\ref{f3theta0q}), (\ref{f3theta0g})
hold for the nucleon as 
\begin{eqnarray}
 \Delta^\mu\bar{u}(p',S')u(p,S) m_N\overline{c}^{(N)}_q(t)&=&\langle N(p',S')|\bar{q}ig F^{\mu \nu}\gamma_\nu q|N(p, S)\rangle\ ,
%\;\;\;\;\left(\mbox{leading to,     } \;\;\; =f_3(t) \right)\ ,
\nonumber\\
\Delta^\mu \bar{u}(p',S')u(p,S) m_N\overline{c}_g^{(N)}(t) &=&-\sum_q\Delta^\mu\bar{u}(p',S')u(p,S) m_N\overline{c}^{(N)}_q(t)
\nonumber\\
&&
%= -\sum_q\langle p'|\bar{q}ig F^{\mu \nu}\gamma_\nu q|p\rangle
%\nonumber\\
%&&= \langle p'|ig F_a^{\mu \nu}\left(-\sum_q\bar{q}t^a\gamma_\nu  q\right)|p\rangle
=\langle N(p',S')|F_a^{\mu \nu} iD^\rho_{ab}  F^b_{\rho \nu}|N(p, S)\rangle\ .
\end{eqnarray}

We also apply the logic used in (\ref{hw0})-(\ref{hw}) to the nucleon case associated with (\ref{para}):
corresponding to (\ref{hw}), we now obtain
\begin{eqnarray}
\langle N(p',S')|T_{q}^{\mu\nu}|N(p, S)\rangle&\simeq&
\frac{1}{2}\left( \langle x \rangle \overline{P}^\mu\langle N(p',S')|\bar{q}\gamma^{\nu} q
|N(p, S)\rangle+ \langle N(p',S')|\bar{q}gA^\mu\gamma^{\nu} q
|N(p, S)\rangle
\right)\nonumber\\
&+& \left( \mu \leftrightarrow  \nu \right)
\nonumber\\
&=&\bar u(p',S')\left[ \langle x \rangle F_1^q(t) \overline{P}^{(\mu}  \gamma^{\nu )} +\langle x \rangle  F_2^q(t)
  \frac{\overline{P}^{(\mu} i\sigma^{\nu ) \alpha}\Delta_\alpha}{2m_N}\right]u(p,S)
\nonumber\\
&+& \frac{1}{2}\langle N(p',S')|\left( \bar{q}gA^\mu\gamma^{\nu}q+ \bar{q}gA^\nu\gamma^{\mu}q\right) 
|N(p, S)\rangle\ ,
\label{hwnucl}
\end{eqnarray}
where $\langle x \rangle$ is the ``average value'' of the quark momentum fraction as in (\ref{averagedef}),
and $F_{1,2}^q (t)$ are the usual Dirac and Pauli form factors for the nucleon.
Comparing this with (\ref{para}), we obtain
\begin{eqnarray}
&&\bar{u}(p',S')\Bigl[A_{q}^{(N)} (t)\frac{\overline P^{\mu}\overline P^{\nu}}{m_N}
 +\bigl(A_{q}^{(N)}  (t)+ B_{q}^{(N)} (t)\bigr)\frac{\overline P^{(\mu}i\sigma^{\nu)\alpha}\Delta_\alpha}{2m_N} \nonumber \\
 && \quad  + D_{q}^{(N)} (t)\frac{\Delta^\mu\Delta^\nu -g^{\mu\nu}\Delta^2}{4m_N} + \bar{c}_{q}^{(N)} (t)m_Ng^{\mu\nu}\Bigr] u(p,S)
\nonumber\\
&\simeq&
\bar u(p',S')\left[ \langle x \rangle F_1^q(t) \frac{\overline P^{\mu}\overline P^{\nu}}{m_N} + \bigl(\langle x \rangle F_1^q(t) +\langle x \rangle  F_2^q(t)\bigr)
  \frac{\overline{P}^{(\mu} i\sigma^{\nu ) \alpha}\Delta_\alpha}{2m_N}\right]u(p,S)
\nonumber\\
&+& \frac{1}{2}\langle N(p',S')|\left( \bar{q}gA^\mu\gamma^{\nu}q+ \bar{q}gA^\nu\gamma^{\mu}q\right) |N(p, S)\rangle
\ .
\label{paranewnovel}
\end{eqnarray}
The $\mu=\nu=+$ component of this formula leads to
the two relations,
\begin{eqnarray}
 A_q^{(N)} (t) +\eta^2D_{q}^{(N)} (t)
&\simeq&
 \langle x \rangle F_1^q(t)\ , 
\nonumber\\
B_q^{(N)} (t)-\eta^2D_{q}^{(N)} (t)&\simeq&
 \langle x \rangle F_2^q(t)\ ,
\label{paranovelr}
\end{eqnarray}
and, similarly, taking the component with $\mu=\perp$ and $\nu=+$
and combining with (\ref{paranovelr}), we find
\begin{eqnarray}
-\frac{\eta\Delta_\perp^\mu\bar{u}(p',S')u(p,S)}{m_N} D_{q}^{(N)}(t)
&\simeq&\langle  N(p',S')|\bar{q}gA_\perp^\mu\Slash{n}q |N(p, S)\rangle
\nonumber\\
&&\!\!\!\!\!\!\!\!\!\!\!\!\!\!\!\!\!\!\!\!\!\!\!\!\!\!\!\!\!\!\!\!
=\frac{1}{2}\int_{-\infty}^\infty d\lambda\ {\rm sgn} (\lambda) 
n_\alpha\langle  N(p',S')| gF_a^{\mu \alpha}(\lambda n)
\bar{q}(0)t^a\Slash{n} q(0)
|N(p, S)\rangle
\ ,
\end{eqnarray}
where (\ref{lcgglue}) is substituted for the gluon field.
This expression suggests that $D_q^{(N)}(t)$ corresponds to an integral of
the twist-three quark-gluon correlation, in particular, 
the correlation between the quark color-current density and the gluon field strength; the flavor singlet 
part can be re-expressed similarly as (\ref{gluoncorrresult}).
These results indicate, for the behaviors of the form factors at large $t$, that
$A_q^{(N)}(t)$ should obey the quark counting, which is same as the vector form factor $F_1^q(t)$,
while $D_q^{(N)}(t)$ should receive an additional $1/t$ suppression due to the counting for the three-body Fock state.
Namely, because $F_1^q(t\to \infty) \sim 1/t^2$ as in \cite{Brodsky:1973kr,Brodsky:1974vy,Matveev:1973ra}, we expect the large $t$ behavior to be
\begin{eqnarray}
A_q^{(N)}(t) \sim \frac{1}{t^2}\ , \;\;\:\;\;\;\; D_q^{(N)}(t) \sim \frac{1}{t^3}\ .
\end{eqnarray}

\section{Conclusions}

We have discussed the QCD constraints on the gravitational form factors 
for a spin-0 hadron. Our results are based not only on the gauge-invariant local operator manipulations,
but also on an approach with light-cone gauge fixing which is linked with the partonic interpretations 
in the infinite momentum frame; for a massless Nambu-Goldstone boson, the infinite momentum frame is likely to be 
an appropriate frame of reference, where the roles of the physical degrees of freedom can be interpreted straightforwardly.
Our results for the explicit operator forms for each Lorentz structure of 
the quark contribution to the form factors indicate that the form factor corresponding to the D term as well as  
the form factor associated with the metric tensor $g^{\mu \nu}$ is due to the higher twist multipartonic correlation effects. 
Further development of the present approach and the applications of the present results will be discussed elsewhere.

\begin{acknowledgments}
I thank 
S.~Kumano and Q.-T. Song
for stimulating discussions.
\end{acknowledgments}

\bibliography{../bibdata/bibdatabase}% Produces the bibliography via BibTeX.

\end{document}